# Why Aβ42 Is Much More Toxic Than Aβ40

J. C. Phillips

Dept. of Physics and Astronomy, Rutgers University, Piscataway, N. J., 08854

Keywords:  protein; function; amino acid; sequence; evolutionary; criticality; self-organized

### Abstract

Amyloid precursor with 770 amino acids dimerizes and aggregates, as do its c terminal 99 amino acids and amyloid 40,42 amino acids fragments.  The titled question has been discussed extensively, and here it is addressed further using thermodynamic scaling theory to analyze mutational trends in structural factors and kinetics.  Special attention is given to Family Alzheimer's Disease mutations outside amyloid 42.  The scaling analysis is connected to extensive docking simulations which included membranes, thereby confirming their results and extending them to Amyloid precursor.

## 1.    Introduction

Alzheimer's disease, the most fatal neurodegenerative disorder, is caused by amyloid-β (Aβ) amyloidogenesis.  Aβ is predominantly a 40 or 42 fragment cleaved from the 770 amino acid A4 precursor protein, which begins at A4 672D (renumbered as 1 in Aβ).  This primary A4 cleavage was explained in [1] as the position of a hydrophilic extremum at 672D in the KD hydropathic profile (Fig. 4 there, partially reproduced below as Fig. 2, for the reader's convenience).  The c-terminal cleavage corresponds to a very strong and flat hydrophobic maximum at 40-42, associated in dilute solution with a large conspicuous uninterrupted hydrophobic patch which covers ~ 25% of the surface [2]. The hydropathic differences between Aβ 40 and 42 are small, but  Aβ  displays  a  dramatically  increased  propensity  to  form  amyloid  fibrils  *in  vitro* [3]. Considering the complexities of both sample preparation and measurement, recent progress in determining the structure of Aβ 42 fibrils [3,4] is remarkable (reviewed in [5]). The dimerized fibril is composed of two parts, with residues 15-42 forming a double-horseshoe-like cross-beta-sheet entity with maximally buried hydrophobic side chains. Residues 1-14 are partially ordered in a looser beta-strand conformation which adjusts reversed and staggered interdimer misfit.

Much study has been devoted to analyzing the molecular origin of the toxicity differences between Aβ 40 and 42 [6-16]. Fibril growth into protofibrils may depend critically on secondary nucleation steps such as amyloid monomer aggregation that is catalysed on the surface of the side of the fibril [16]. The aggregating monomers may be Aβ42 + a transmembrane tail extending 4-12 amino acids of A4 beyond 42A up to 54KKK56. There are many familial mutations in Aβ or adjacent to its hydrophilic 672 n-terminal, most of which are related to its salt bridge and beta strand side turn packing [3]. Three familial (FAD) mutations in the tail region occur at Val 46: Phe, Ile, Gl [14].

## 2. Results

We wish to analyze the systematics of all seven nearby Val46 mutations of [14]: FAD Phe, Ile, Gln, and the artificial non-FAD Cys, Ser, Glu, Lys. The data in percents distinguish between Aβ1-42/totalAβ and Aβx-42/totalAβ. Here Aβx includes heterogeneous Ile, Val, Arg Glu, and benzenoid Phe, mutations at Asp site 1, while totalAβ means the sum of Aβ40,42 and 43, as shown in columns 1,2 in Table 1 of [14]. To gain insight into the mechanisms selecting Aβ42, we use four scales: (1) long range hydropathic [15], short range hydropathic [16], (3) beta strand buried [17] and (4) alpha helical buried [17] propensities. Another factor influencing the fraction of Aβ42 is the shape of the 46 mutation. Among the mutants only Phe contains a bulky benzoid side group [18]. We therefore correlated the results of [14] to four sets of mutants: FAD only and FAD + non-Fad, with and without Phe. This generates a table of 32 correlations, mostly weak. The complete results are included as supplementary information in Archiv, while here we discuss only highlights.

The strongest correlations involve ABx-42 (heterogeneous n-terminal). They also exclude Phe, and thus support the conclusion of [18]: the bulky benzenoid Phe side group makes the transmembrane handle 43-46 more brittle, and facilitates cleavage at 42, explains why F46 produces the largest yields for both Aβ1-42/totalAβ and Aβx-42/totalAβ. Excluding Phe leaves only three points: WT Val, Ile and Gln, but the correlations are strong. They range from 0.82 (beta strand and alpha helical buried) to 0.98 (long range hydropathicity). A nice touch is that the latter is markedly better than short range hydropathicity (0.90), and both are better than secondary strand or helical propensities.

All of these correlations are drastically reduced for Aβ1-42/totalAβ, by 0.3-0.5. Either restriction of amyloid β to n-terminal Asp site 1 is artificial, or there is some unknown factor affecting the restriction of the data to Asp 1. The correlations of all site 46 mutants (FAD + non-FAD, except Phe) is strongest for buried alpha helical propensity with Aβx-42/totalAβ (0.87), and again weaker for Aβ1-42/totalAβ (0.75).

We now turn to the broader perspective of the A4 precursor protein profile (Fig. 1). The striking feature here is the very strong hydrophobic peak centered near 740, which is shown in detail in Fig. 2. Cleavages at 1 (Aβ numbering) and 40,42 are associated with hydrophilic and hydrophobic extrema for W = 19-23, but outside this range the correspondence deteriorates; thus the cleavage success in hydropathic profiles is associated with the membrane wavelength W ~ 21.

The FAD mutations at Val46 lie near the 40-42 cleavage, but outside Aβ itself. Can we learn something from secondary (α and β) buried propensities? The natural value for W here is 13, because this is the beta strand horseshoe length [3,4]; see also Fig. 2 of [1]. A curious feature of the eight combined mutations FAD (including Phe) + artificial non-FAD is that they correlate well (0.74) with buried β strand propensity. This could be related to the Val46 extremum shown in Fig. 3. Surprisingly there is little correlation of Aβ1-42/totalAβ and Aβx-42/totalAβ with FAD only with or without Phe. Thus, in spite of the successful identification of Val46 as a candidate for epsilon cleavage, the Val46 mutation route to toxic Aβx-42 or Aβ1-42 lies either through benzenoid Phe or reduced hydrophobicity.

## 3. Discussion

Dimerization occurs not only in 40 amino acid Aβ fragments of A4, but also in A4 itself [19], as well as C99, the c-terminal 99 (672-770) amino acid fragment. Moreover, the 43TVIV46 sites adjacent to Aβ42 are critical determinants of C99 dimerization [20]. The common factor in all three cases is the very strong hydrophobic peak in Fig. 1. There have been many docking studies of C99 involving membrane interactions. Indeed [21] found that "C99 was anchored at the boundary of the lipid raft relying on the conserved hydrophobic motif of V(710)xxA(713)xxxV(717)xxxV(721) ". This matches exactly the W = 21 hydroprofile features shown in Figs. 1 and 2. Of course 710 is the hydrophobic peak, while the 721 cutoff at 185 is level [22,23] with both the second-highest 120 and 180 n-terminal peaks in Fig. 1. Thus hydroprofiling with the membrane wavelength W = 21 has succeeded in extending the docking calculations [21] for a ~ 100 amino acid fragment to the entire ~ 800 amino acid A4.

This is a striking example of the importance of level sets in identifying hydrodynamic aspects of molecular dynamics simulations ([22] and Sec. 7 of [23]) that are thought to be technically inaccessible to simulations [24,25]. There are many examples [23] of level set synchronization, for instance, HPV vaccine, aspirin (COX-2), myoglobin and neuroglobin, macroglobulin, enzyme prostaglandin-H2 D-isomerase (P2GDS), and others more recent [26]. The present match to a sophisticated docking calculation is suggestive of the potential gains by connecting thermodynamic scaling to molecular dynamics simulations.

While it is generally agreed that unmutated Aβ42 is more toxic than unmutated Aβ40, an explanation for its presumably enhanced Aβ42/Aβ40 aggregation rate [25] has remained elusive [8,27-31]. From Fig. 2, the area under the membrane W = 21 KD curve from the critical I41 (702) level of 185 up to Ala42 is 50% greater than that area up to Val40. This difference could easily account for both the Aβ42/Aβ40 differences in gamma secretase cleavage rates and aggregation rates, with features characteristic of Aβ42/Aβ40, C99 and even A4.

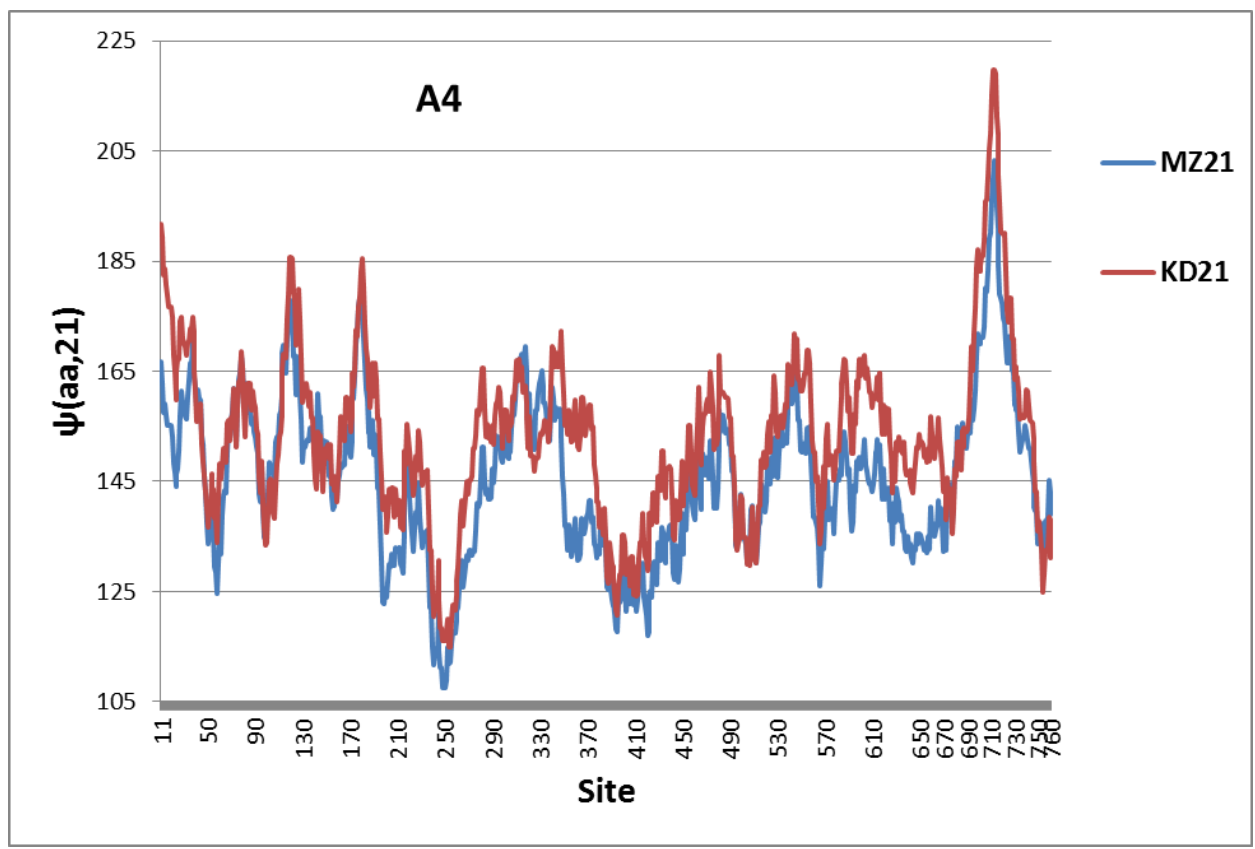

Fig. 1. Hydropathic profiles of the precursor protein A4, using the membrane wave length W = 21, with the short(KD,[16])- and long(MZ,[15])-range hydropathic scales. For most of the protein the oscillations fall in a normal range 125-185 centered on hydroneutral 155. The 80 amino acid c-terminal hydrophobic peak reaches as high as 215, which is twice as high relative to hydroneutral as the upper limit of the normal range. Amyloid β is formed from the left side of this peak, as shown in detail in Fig. 2.

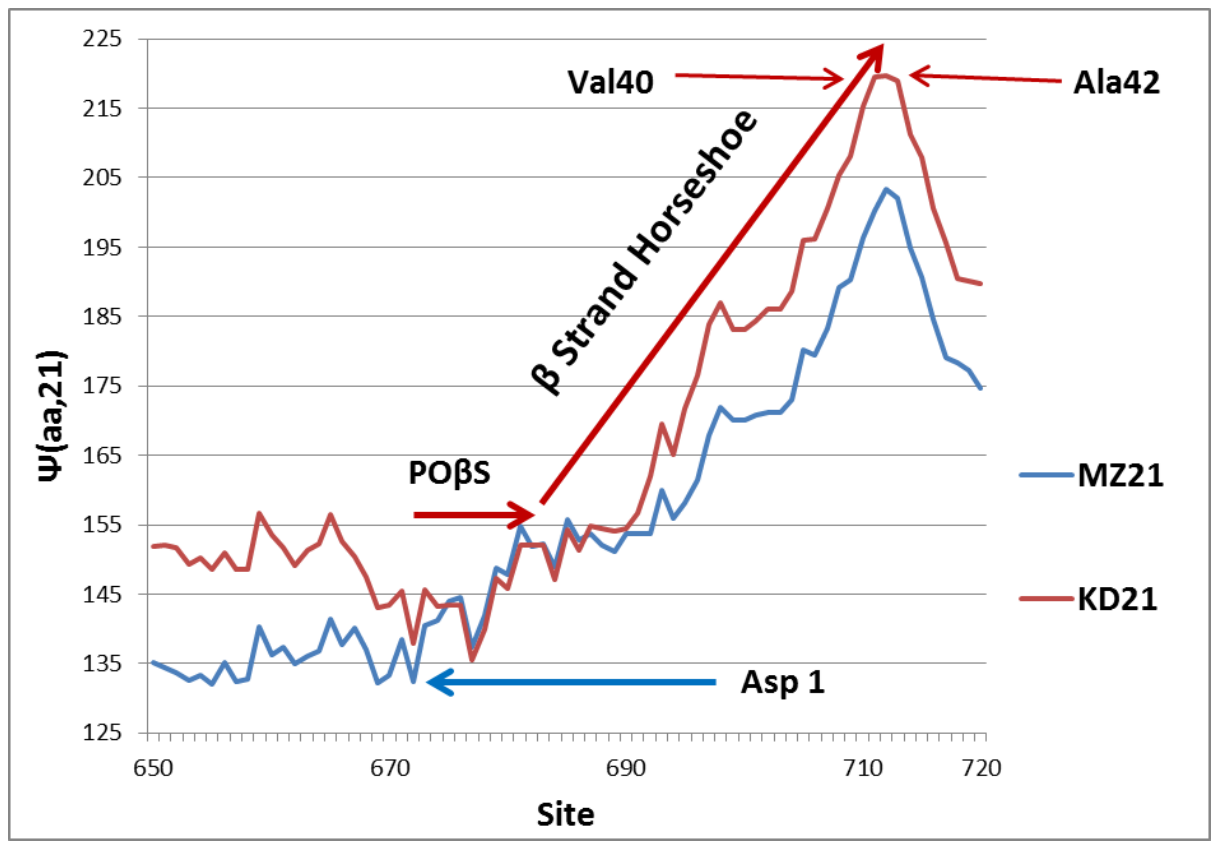

Fig. 2. With the membrane wave length W = 21 there is an excellent correpondence between the details of the hydropathic profiles and the dimer fibril NMR beta strand structure [3,4]. The n-terminal cleavage at 672 occurs at a hydrophiic minimum where the Partially Orderded β Strand structure begins. The β Strand Horseshoe is strongly amphilic, and reaches a flat hydrophobic peak at 40,42.

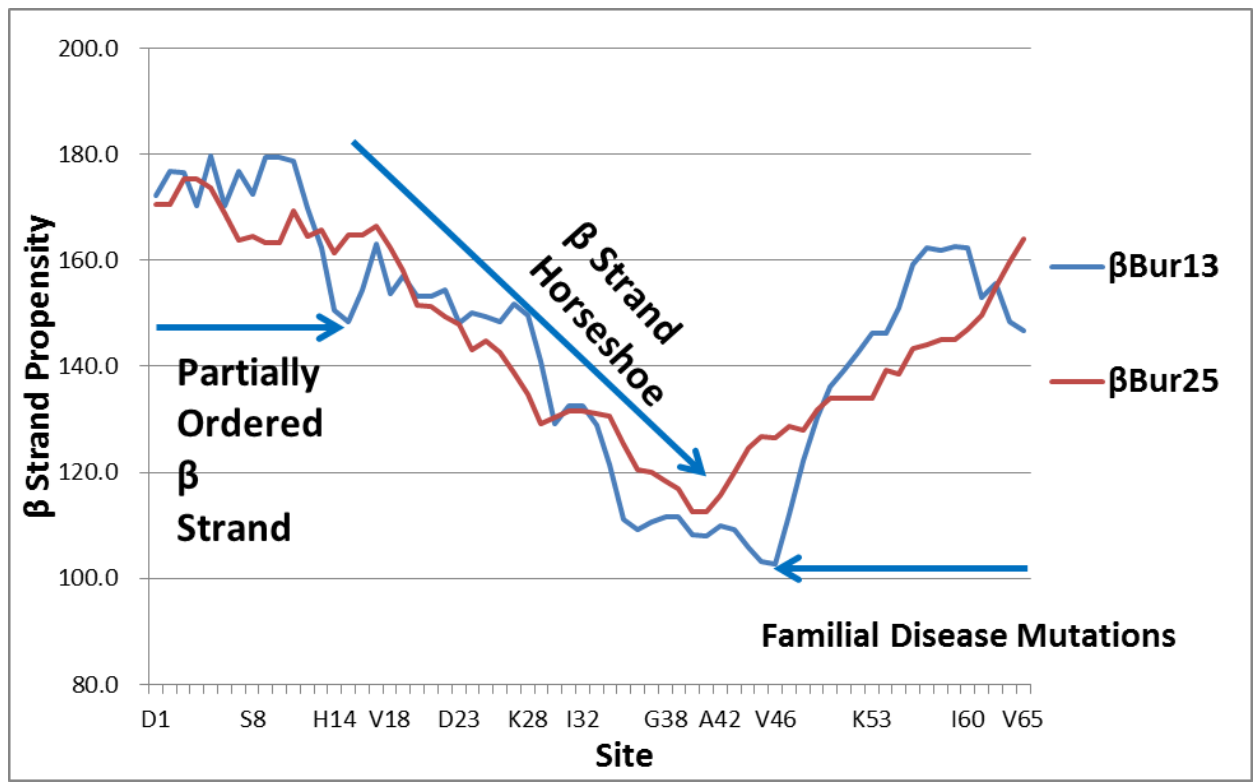

Fig. 3.  The buried β strand propensity waves correlate with the known beta strand structure [3,4] and with the natural choice W = 13 identifies the Val46 FAD mutations with a minimum.

| | | C | S | E | K | I | G | WT | F |
|---|---|---|---|---|---|---|---|---|---|
| 1 | Aβx- 42 | 15 | 17 | -25 | 9 | 26 | 45 | 14 | 52 |
| 2 | Aβ1- 42 | 13 | 15 | -17 | 8 | 26 | 25 | 14 | 42 |
| 3 | MZ | -8 | 138 | 144 | 169 | 16 | 82 | 0 | 20 |
| 4 | KD | 34 | 99 | 152 | 160 | -6 | 91 | 0 | 28 |
| 5 | β strand | -7 | -46 | -121 | -122 | 12 | -42 | 0 | 1 |
| 6 | α prop | 0.06 | 0.15 | -0.48 | -0.26 | -0.13 | 0.47 | | -0.09 |

| | All eight | FAD only | 7 Without F | 3 Without F |
|---|---|---|---|---|
| (1,3) | -0.46 | 0.59 | -0.38 | 0.98 |
| (1,4) | -0.50 | 0.66 | -0.45 | 0.90 |
| (1,5) | 0.64 | -0.43 | 0.60 | -0.82 |
| (1,6) | 0.68 | 0.31 | 0.87 | 0.82 |
| (2,3) | -0.54 | 0.12 | -0.49 | 0.59 |
| (2,4) | -0.63 | 0.18 | -0.62 | 0.38 |
| (2,5) | 0.74 | 0.10 | 0.73 | -0.23 |
| (2,6) | 0.54 | -0.22 | 0.75 | 0.24 |

Table I. Details of the analysis of the data of [14], which are the first two lines of the upper table. The four factors used are listed in lines 3-6, relative to their wild type (WT) values. The lower table gives the Pearson correlations of the data with the factors. The significance of the correlations is discussed in the text.